\documentclass[fleqn,usenatbib]{mnras}

\usepackage{amsmath}	

\usepackage{txfonts}

\usepackage[T1]{fontenc}
\usepackage{ae,aecompl}

\usepackage{hyperref}	
\hypersetup{colorlinks=true,linkcolor=blue,citecolor=blue,filecolor=blue,urlcolor=blue}
\usepackage{graphicx}	
\usepackage{amssymb}	
\usepackage{color}

\newcommand{\fesc}{\ifmmode{f_{\rm esc}}\else{$f_{\rm esc}$}\fi}
\newcommand{\fescs}{\ifmmode{f_{\rm esc}^\star}\else{$f_{\rm esc}^\star$}\fi}
\newcommand{\kms}{\ifmmode{{\;\rm km~s^{-1}}}\else{km~s$^{-1}$}\fi}
\newcommand{\fgas}{\ifmmode{{f_{\rm gas}}}\else{$f_{\rm gas}$}\fi}
\newcommand{\cubecm}{\ifmmode{{\rm cm^{-3}}}\else{cm$^{-3}$}\fi}
\newcommand{\ztwo}{\ifmmode{{\rm [Z_2/H]}}\else{[Z$_2$/H]}\fi}
\newcommand{\zthree}{\ifmmode{{\rm [Z_3/H]}}\else{[Z$_3$/H]}\fi}
\newcommand{\lsim}{\lower0.3em\hbox{$\,\buildrel <\over\sim\,$}}
\newcommand{\gsim}{\lower0.3em\hbox{$\,\buildrel >\over\sim\,$}}

\newcommand{\flux}{erg s$^{-1}$ cm$^{-2}$ Hz$^{-1}$}
\newcommand{\emis}{erg s$^{-1}$ cm$^{-2}$ Hz$^{-1}$ sr$^{-1}$}
\newcommand{\sfr}{\ifmmode{\textrm{M}_\odot \,\textrm{yr}^{-1} \,\textrm{Mpc}^{-3}}\else{M$_\odot$ yr$^{-1}$ Mpc$^{-3}$}\fi}
\newcommand{\hsfr}{\ifmmode{\textrm{M}_\odot\, \textrm{yr}^{-1}}\else{M$_\odot$ yr$^{-1}$}\fi}

\newcommand{\eavg}{\ifmmode{\langle E_\gamma \rangle}\else{$\langle E_\gamma \rangle$}\fi}

\newcommand{\Ms}{\ifmmode{M_\odot}\else{$M_\odot$}\fi}
\newcommand{\vrms}{\ifmmode{v_{\rm rms}}\else{$v_{\rm rms}$}\fi}

\newcommand{\hh}{H$_2$}

\newcommand{\tvir}{\ifmmode{T_{\rm{vir}}}\else{$T_{\rm{vir}}$}\fi}
\newcommand{\mvir}{\ifmmode{M_{\rm{vir}}}\else{$M_{\rm{vir}}$}\fi}
\newcommand{\rvir}{\ifmmode{r_{\rm{vir}}}\else{$r_{\rm{vir}}$}\fi}

\newcommand{\jj}{\ifmmode{J_{21}}\else{$J_{21}$}\fi}
\newcommand{\flw}{\ifmmode{F_{LW}}\else{$F_{LW}$}\fi}
\newcommand{\kph}{\ifmmode{k_{\rm ph}}\else{$k_{\rm ph}$}\fi}

\newcommand{\msun}{{\rm\,M_\odot}} 

\newcommand{\zsun}{\ifmmode{\rm\,Z_\odot}\else{$\rm\,Z_\odot$}\fi}

\newcommand{\hii}{H {\sc ii}}

\newcommand{\nhi}{\ifmmode{N_{\rm HI}}\else{$N_{\rm HI}$}\fi}

\def\eps@scaling{1.0}%
\newcommand\epsscale[1]{\gdef\eps@scaling{#1}}%
\newcommand\plotone[1]{%
 \centering 
 \leavevmode 
 \includegraphics[width={\eps@scaling\columnwidth}]{#1}%
}%
\newcommand\plottwo[2]{%
 \centering 
 \includegraphics[width={\eps@scaling\columnwidth}]{#1}%
 \hfil 
 \includegraphics[width={\eps@scaling\columnwidth}]{#2}%
}%

\title[Reionisation with First Galaxies]{Extending Semi-numeric Reionisation Models to the First
  Stars and Galaxies}

\author[Koh \& Wise]{
Daegene Koh,$^{1,2}$\thanks{E-mail: kdaegene@stanford.edu}
John H. Wise,$^{1}$\thanks{E-mail: jwise@gatech.edu}
\\
$^{1}$Center for Relativistic Astrophysics, Georgia Institute of
Technology, 837 State Street, Atlanta, GA 30332, USA\\
$^{2}$Kavli Institue for Particle Astrophysics and Cosmology, Stanford
University, Menlo Park, CA 94025, USA\\
}

\pubyear{2016}

\begin{document}
\label{firstpage}
\pagerange{\pageref{firstpage}--\pageref{lastpage}}
\maketitle

\begin{abstract}

Semi-numeric methods have made it possible to efficiently model the epoch of reionisation (EoR). While most implementations involve a reduction to a simple three-parameter model, we introduce a new mass-dependent ionising efficiency parameter that folds in physical parameters that are constrained by the latest numerical simulations. This new parameterization enables the effective modeling of a broad range of host halo masses containing ionising sources, extending from the smallest Population III host halos with $M \sim 10^6 M_\odot$, which are often ignored, to the rarest cosmic peaks with $M \sim 10^{12} M_\odot$ during EoR. We compare the resulting ionising histories with a typical three-parameter model and also compare with the latest constraints from the \textit{Planck} mission. Our model results in a optical depth due to Thomson scattering, $\tau_{\mathrm{e}}$ = 0.057, that is consistent with \textit{Planck}. The largest difference in our model is shown in the resulting bubble size distributions which peak at lower characteristic sizes and are broadened. We also consider the uncertainties of the various physical parameters and comparing the resulting ionising histories broadly disfavors a small contribution from galaxies. As the smallest haloes cease a meaningful contribution to the ionising photon budget after $z = 10$, implying they play a role in determining the start of EoR and little else.
\end{abstract}

\begin{keywords}
  dark ages, reionization, first stars -- galaxies:high-redshift -- early Universe -- stars: Population III -- cosmology:theory
\end{keywords}



\section{Introduction}

Models of the Epoch of Reionisation (EoR) have been extensively improved over the years as tighter observational constraints are provided. This particular phase transition of the universe can provide a number of insights into the details of the beginnings of structure formation \citep[e.g.][]{Robertson2010}. 

The biggest current observational constraints in modeling EoR comes from the Thomson scattering optical depth, $\tau_{\mathrm{e}}$, to the cosmic microwave background (CMB). Improvements to the measurement have progressively driven down this particular value, where the latest results estimate $\tau_{\mathrm{e}}$ = 0.0596 $\pm$ 0.0089 corresponding to a redshift of instantaneous reionsation of z = $8.0_{-1.1}^{+0.9}$ \citep{Planck2016}. Next, the transmission fraction of quasar light through the intergalactic medium shows that the EoR ended by z $\sim$ 6 \citep{Gunn1965, Fan2006}, though there are some recent observations implying that it may not have been completed until z $\sim$ 5.6 \citep{Mesinger2010}. Interestingly, it was previously expected that quasars alone could not produce the needed number of ionising photons to complete reionisation as earlier constraints from Wilkinson Microwave Anisotropy Probe (WMAP) introduced a need for high-redshift sources \citep[e.g.][]{Willott2010, Grissom2013}. However, with the updated Planck results requiring a later start to reionisation, quasars have come back in recent models, in some of which they are the only sources \citep{Madau2015}. 

Theoretical efforts in the modeling EoR has ranged from relative simple analytic models \citep{Madau1999, Kuhlen2012, Alvarez2012} all the way to high-resolution numerical simulations with various detailed star and galaxy formation prescriptions that include self-consistent ray tracing \citep{Iliev2007,Trac2011}. Reionisation necessitates a large number of approaches due to the wide range of scales involved in the process. Moreover, a full numerical solution would require parsec scale resolution to correctly follow sources and feedback in, at minimum, a 100 comoving $\rm{Mpc}^3$ box to get convergent histories \citep{Iliev2014}. Such simulations would be an enormous computational cost.

Semi-numeric models are thus an attractive alternative. Such models can accurately generate full three-dimensional density, velocity, and ionisation fields without the need to follow the underlying physics \citep{Mesinger2007, Zahn2007}. These models make the following fundamental assumption that overdense regions drive the ionisation process. With this assumption, one asserts that if the number of available photons exceeds the number of baryons in a cell, the cell must be ionised. This simple model provides a powerful tool that compares favorably with high-resolution radiative transfer numerical simulations \citep{Zahn2011}.

Within the numerous models, there have been many efforts to understand the role of the various potential sources in the reionisation process. Typical models only consider galaxies hosted by atomic-cooling halos above $T_{\rm{vir}}$ $\sim 10^4$ K. However, an often neglected source is mini-halos with $M < 10^8 \msun$ containing massive, metal-free Pop III stars. Pop III stars have been studied extensively over the past decade detailing their formation \citep{Abel2002, Turk2009, Greif2011}, their spectral properties \citep{Tumlinson2000, Schaerer2002}, and their final fates \citep{Woosley1995, Heger2002, Heger2010}. Of more interest to the EoR, these massive stars also produce extended \hii{} regions in their immediate vicinity spanning 1-3 kpc \citep{Kitayama2004, Whalen2004, Alvarez2006, Abel2007}. These \hii{} regions will then grow out further as mini-halos merge together to form the first galaxies providing additional ionising flux. \citet{Ahn2012} used a sub-grid model to populate mini-halos in a 114 Mpc $h^{-1}$ simulation and showed their addition had a significant effect in determining the onset of reionisation. Furthermore, \citet{Wise2014} calculated the escape fraction of ionising photons in a 1 comoving Mpc radiation hydrodynamics simulation showing that mini-halos can contribute up to 30 percent of the ionising photon budget. 

Given the extensive volume of data available from large volume high-resolution simulations at high redshifts, we can take simulated physical properties of ionising sources, such as the photon escape fraction and star formation efficiency. In this work, we take these calculated properties to create a new parameterization extending existing semi-numeric models to include the effects of mini-halos. 

In the immediately following Section \ref{sec:Methods}, we introduce our new parameterization. In Section \ref{sec:Results}, we compare the ionisation histories produced from our model and show the resulting bubble size distributions. Finally, in section \ref{sec:disc}, we provide a short discussion and summarize our results.

\section{Methods}
\label{sec:Methods}

\subsection{Simulating Reionisation}

Our treatment involves use of the semi-numerical reionisation simulation code {\sc{21cmFAST}} \citep{Mesinger2011}. In this code, the ionisation field is generated following an excursion-set approach \citep{Furlanetto2004}. Namely, a cell is considered to be ionised when 

\begin{equation}
f_{\rm{coll}} (x, M_{\rm{min}}, R, z) \geq \zeta^{-1} 
\end{equation}
where $\zeta$ is the ionisation efficiency, and $f_{\rm{coll}}$ is the fraction of collapsed mass inside a region of size $R$ in halos whose mass is greater than $M_{\rm{min}}$ \citep{Mesinger2007, Zahn2007}. This value $R$ is iterated from $R_{\rm{max}}$, which is typically taken to be the maximum horizon of ioinising photons, or the effective mean free path down to the length of a single cell. These three parameters then fully determine the ionisation state at any given redshift. Our simulations are run on a box with a 100 comoving Mpc side length using $2048^3$ cells down-sampled to $1024^3$ cells to generate the ionisation field. The main contribution in this work is our detailed treatment of the parameter $\zeta$ which is outlined in the following sections.

\subsection{Calculating the Ionising Efficiency}

In previous treatments, $\zeta$ typically represents a homogeneous ionising efficiency factor for all star-forming galaxies in any environment. A typical parameterization is provided in \citet{Greig2016} as 

\begin{equation}
\zeta = 30\ \left(\frac{\fesc}{0.2}\right) \left(\frac{f_*}{0.03}\right) \left(\frac{f_\mathrm{b}}{\Omega_\mathrm{b}/\Omega_\mathrm{m}}\right) \left(\frac{N_{\mathrm{\gamma/b}}}{4000}\right) \left(\frac{1.5}{1+n_{rec}}\right)
\end{equation}
where $\fesc$ is the fraction of ionising photons escaping into the intergalactic medium (IGM), $f_*$ is the fraction galactic gas in stars, $f_\mathrm{b}$ is the baryon fraction inside haloes hosting galaxies in units of the cosmic baryon fraction, $N_{\rm{\gamma/b}}$ is the number of ionising photons per baryon in stars, and $n_{\rm{rec}}$ is the average number of recombinations per baryon in the IGM. These values are all assumed to be mass- and redshift-independent to produce a single $\zeta$ value. We take this model to be the Fiducial case and compare our new parameterization against it in Sec \ref{sec:Results}. 

The main improvement of this work is to model $\zeta$ as a function of the host halo mass at a given redshift. This class of parameterizations has been initially explored by \citet{Furlanetto2005} but they only considered a simple power-law function setting $\zeta \sim m^{\alpha}$ for various values of $\alpha$. In our work, we consider a more extensive dependence on the host halo mass that is better physically motivated. This allows us to incorporate the distribution of ionising efficiencies at different masses as well including the contribution of mini-halos to the photon budget for reionisation.

In particular, $\zeta$ has been parameterized as follows. 

\begin{equation}
\label{eq:zeta}
\zeta (M_{h})= \begin{cases} 
\zeta_{0,3}\  \fesc \ f_*\ N_{3,\mathrm{\gamma / b}}\ f_{\rm{d}}  \frac{1}{1+n_{\rm{rec}}}  \\
\qquad \qquad \qquad \qquad \qquad \qquad \rm{for} \ M_{\rm{min}} \leq M_{\mathrm{vir}}  < M_{\rm{filter}} \\
\zeta_{0,2}\ \fesc (M_{\mathrm{h}})\ f_*(M_{\mathrm{h}})\ N_{2,\rm{\gamma / b}}\ f_{\rm{d}}(M_{\mathrm{h}}) \frac{1}{1+n_{\rm{rec}}} \\\qquad \qquad \qquad \qquad \qquad \qquad \rm{for} \ M_{\mathrm{vir}} \geq M_{\rm{filter}}
\end{cases}
\end{equation}

where $N_{3,\rm{\gamma / b}}$ is $N_{\rm{\gamma/b}}$ for Pop III stars, $f_{\rm{d}}$ is the effective duty cycle, $N_{2,\rm{\gamma / b}}$ is $N_{\rm{\gamma/b}}$ for galaxies, and $n_{\rm{rec}}$ is the number of recombinations. Lastly, $\zeta_{0,3}$ and $\zeta_{0,2}$ are constants calibrated to the desired reionisation history. In this work, we take these values to be 1.5 and 2.6 respectively. 

Furthermore, the domain of $\zeta$ is characterized by two different masses. First is $M_{\rm{min}}$, which is the minimum mass of mini-halos that is required to collapse to form Pop III stars. This mass is determined by the strength of the soft \hh\ photodissociating Lyman-Werner (LW) flux by

\begin{equation}
\label{eq:mmin}
M_{\rm{min}} (\rm{F_{LW}}) = 1.25\ \rm{x}\ 10^5  + 8.7\ \rm{x}\ 10^5 \left(\frac{F_{LW}}{10^{-21}}\right)^{0.47}
\end{equation}
taken from \citet{Machacek2001}, where $\rm{F_{LW}}$ is the strength of the LW background in units of \flux. The magnitude of this flux as a function of redshift is modeled as

\begin{equation}
\mathrm{log}\ J_{21} (z) = A + Bz + Cz^2 + Dz^3 + Ez^4
\end{equation}
where (A,B,C,D,E) = (-2.567, 0.4562, -0.02680, 5.882 x $10^{-4}$, -5.056 x $10^{-6}$) taken from \citet{Wise2012a}. Here $J_{21}$ is the specific intensity in units of \emis. In this fit, the strength of the background peaks at z = 13.765 with a value of $J_{21} = 0.97$ after which galaxies would dominate the contribution. As the actual minimum mass for collapse would be dependent on the exact environment of the halos, we set the LW background to be this maximum value of this fit at subsequent redshifts. At these redshifts, the exact value has minimal impact on the resulting reionisation history because galaxies provide the bulk of the photon budget.

The other relevant characteristic mass is the filtering mass, $M_{\rm{filter}}$, which is the characteristic mass scale 
below which reionisation suppresses gas fraction in low-mass halos \citep{Gnedin2000} given by 
\begin{equation}
M_{filter}^{2/3} = \frac{3}{a}\int^{a}_{0} da' M_{J}^{2/3} (a') \left[ 1 - \left(\frac{a'}{a}\right)^{1/2}\right]
\end{equation}
where $M_{J}$ is the Jeans mass and $a$ is the cosmological scale factor. Halos above the filtering mass have the sufficient gas content to produce stars. While typical star formation is suppressed in halos below this threshold, these halos can still host primordial stars. As we do not explicitly track metal content, we make the simplifying assumption that all halos below this threshold are Pop III host halos. We calculated $M_{\rm{filter}}$ from the simulations of \citet{Wise2012a} and created a polynomial fit as a function of redshift for computational ease given by 

\begin{equation}
\mathrm{log}\ M_{\rm{filter}} (z) = A + Bz + Cz^2 + Dz^3
\end{equation}
where (A,B,C,D) = (9.065, -0.15611, 0.0063, -1.9577 x $10^{-4}$). This $M_{\rm{filter}}$ is then used to as the mass cut-off above which galaxy formation occurs at a given redshift. From $M_{\rm{min}}$ to $M_{\rm{filter}}$, we assume Pop III stars are the dominant contributors, while for $M_\mathrm{vir} > M_{\rm{filter}}$, galaxies dominate. For the rest of this work, we define mini-halos as halos with masses in the range $M_{\rm{min}} \leq M_{\mathrm{vir}}  < M_{\rm{filter}}$ whose dominant ionising source is Pop III stars.

\begin{figure}
	\includegraphics[width=\columnwidth]{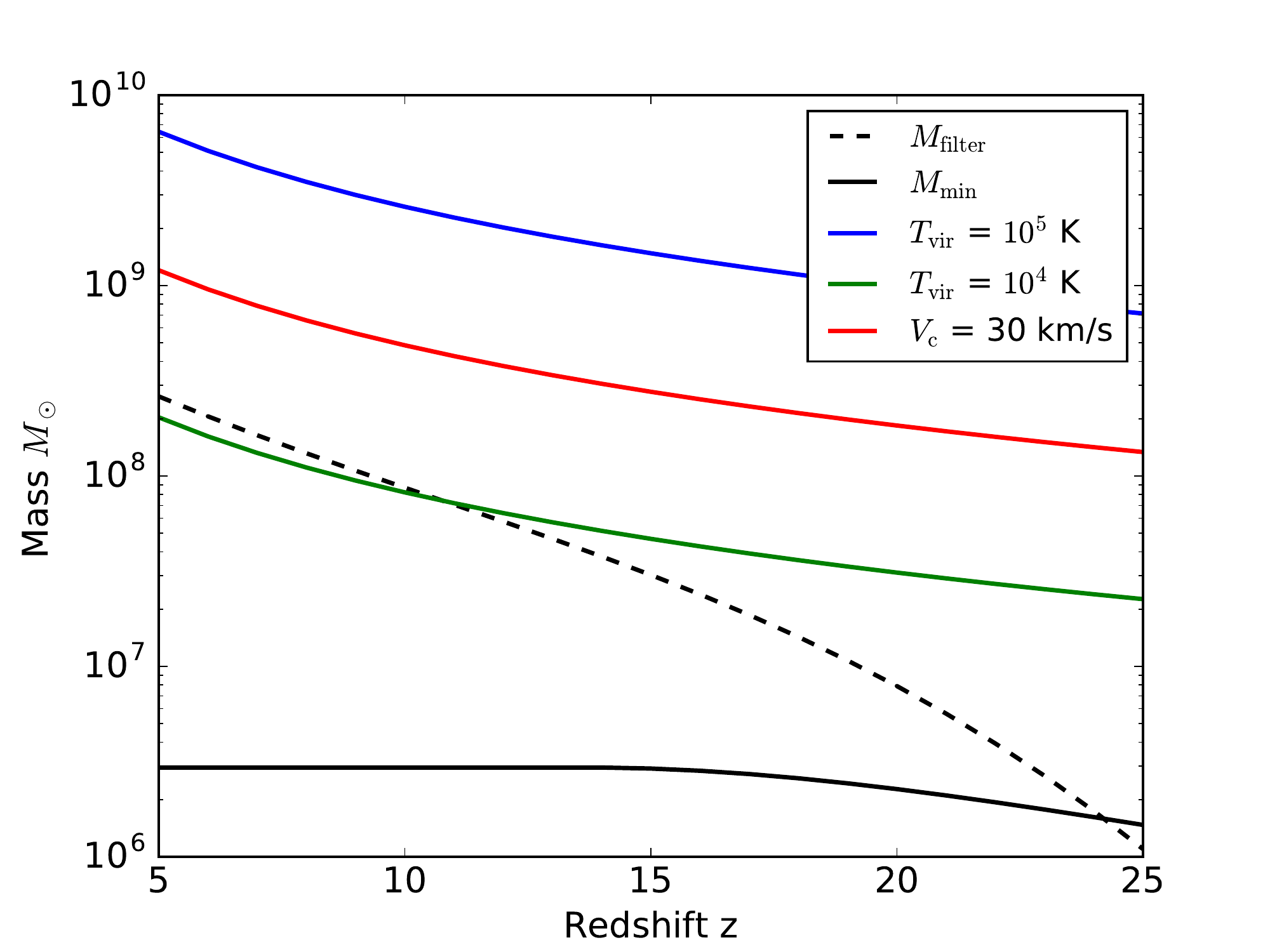}
	\caption{Characteristic masses as a function of redshift. The minimum mass, $M_{\rm{min}}$ (black, solid), is the mass above which ionising sources exist. The filtering mass, $M_{\rm{filter}}$ (black, dashed), is the threshold between Pop III mini-halos and galaxies. That is, any mass range between these two lines will be assumed to be a mini-halo, while any mass range above the dashed line will host galaxies. For comparison, we also show commonly assumed minimum masses corresponding to a virial temperature $T_{\rm{vir}} = 10^4$ K (green), $T_{\rm{vir}} = 10^5$ K (blue), and circular velocity, $V_{\mathrm{c}}$ = 30 km/s (red).}
	\label{fig:masses}
\end{figure}

These characteristic masses are shown in Fig \ref{fig:masses}. When $M_{\rm{filter}}$ > $M_{\rm{min}}$, which happens at z > 24, we set $\zeta$ to be 0 as no galaxies can be formed. We can see that our adopted $M_{\rm{filter}}$ is much less than any of the typically adopted minimum mass values at z > 15, greatly increasing the number of available galaxies to produce ionising photons. 

\subsubsection{Mini-halos}

For mini-halos ($M_{\rm{min}} \leq M_{\mathrm{vir}}  < M_{\rm{filter}}$), the photon contribution is assumed to be entirely from Pop III stars. Given the large uncertainty and lack of observational constraints for the relevant parameters for the first stars, we take each value in the parameterization to be mass- and redshift-independent in the relevant ranges for Pop III stars. Instead, we consider the possible range of values in Sec. \ref{sec:uncertainties}.

First, \fesc\ is the parameter with the largest uncertainty \citep{Alvarez2006}, which we adopt a value of 0.6. We set $f_*$ to be a constant at $100 \msun / 10^6 \msun$ which is a typical ratio found in cosmological simulations of Pop III star formation \citep{Susa2014, Hirano2015}. The number of photons per baryon, $N_{3,\rm{\gamma / b}}$ is largely determined by the surface temperature of the star. We take this value to be 50,000 \citep{Schaerer2002}. Lastly, the duty cycle, $f_{\rm{d}}$, is taken as the fractional star-formation timescale defined as the average lifetime of a Pop III star over the recovery time, to account for the fact that Pop III formation events are bursty. Pop III stars can very efficiently photoevaporate their surroundings and their supernova completely disrupt the host halo \citep{Greif2007, Koh2016}. This results in a significant delay until the subsequent generation of star formation \citep{Jeon2014}. We take this value to be $5\ \rm{ Myr}/30\ \rm{ Myr}$.

\subsubsection{Galaxies}

For halos with $M_\mathrm{h} > M_{\rm{filter}}$, galaxies dominate the photon budget following the death of Pop III stars. For this range of masses, we take a number of fits from cosmological galaxy simulations to calculate $\zeta$. 

The photon escape fraction, \fesc, is modeled using the piece-wise fit below

\begin{equation}
\mathrm{log}\ \fesc (M_{vir}) = \begin{cases} 
-0.51 - 0.039\ \mathrm{log}\ M_{\rm{vir}} &  \mathrm{log}\ M_{\rm{vir}} \geq 8.5\\ 
2.669 - 0.413\ \mathrm{log}\ M_{\rm{vir}} &  7 \leq \mathrm{log}\ M_{\rm{vir}} < 8.5\\
-0.222 &  \mathrm{log}\ M_{\rm{vir}} < 7\\ 
\end{cases}
\end{equation}
taken from \citet{Kimm2014} who used high-resolution zoom-in simulations to construct the fit. This fit takes a nominal value of $\fesc = 0.6$ for halos below log $M_{\rm{vir}}/\msun < 7$, matching our assumed value for mini-halos, with a steep decrease for 7 < log $M_{\rm{vir}}/\msun < 8.5$ and then flattens off for log $M_{\rm{vir}}/\msun > 8.5$ to $\fesc \sim 0.1$. This is consistent with other simulations showing high escape fractions for low mass galaxies \citep{Wise2009, Paardekooper2015}. These values should be taken as estimates of the physical escape fractions, as it does not account for the absorption of photons below the resolution scale; on the other hand, turbulence can enhance the transmission of ionizing photons \citep{Safarzadeh2016}.

\begin{figure}
	\includegraphics[width=\columnwidth]{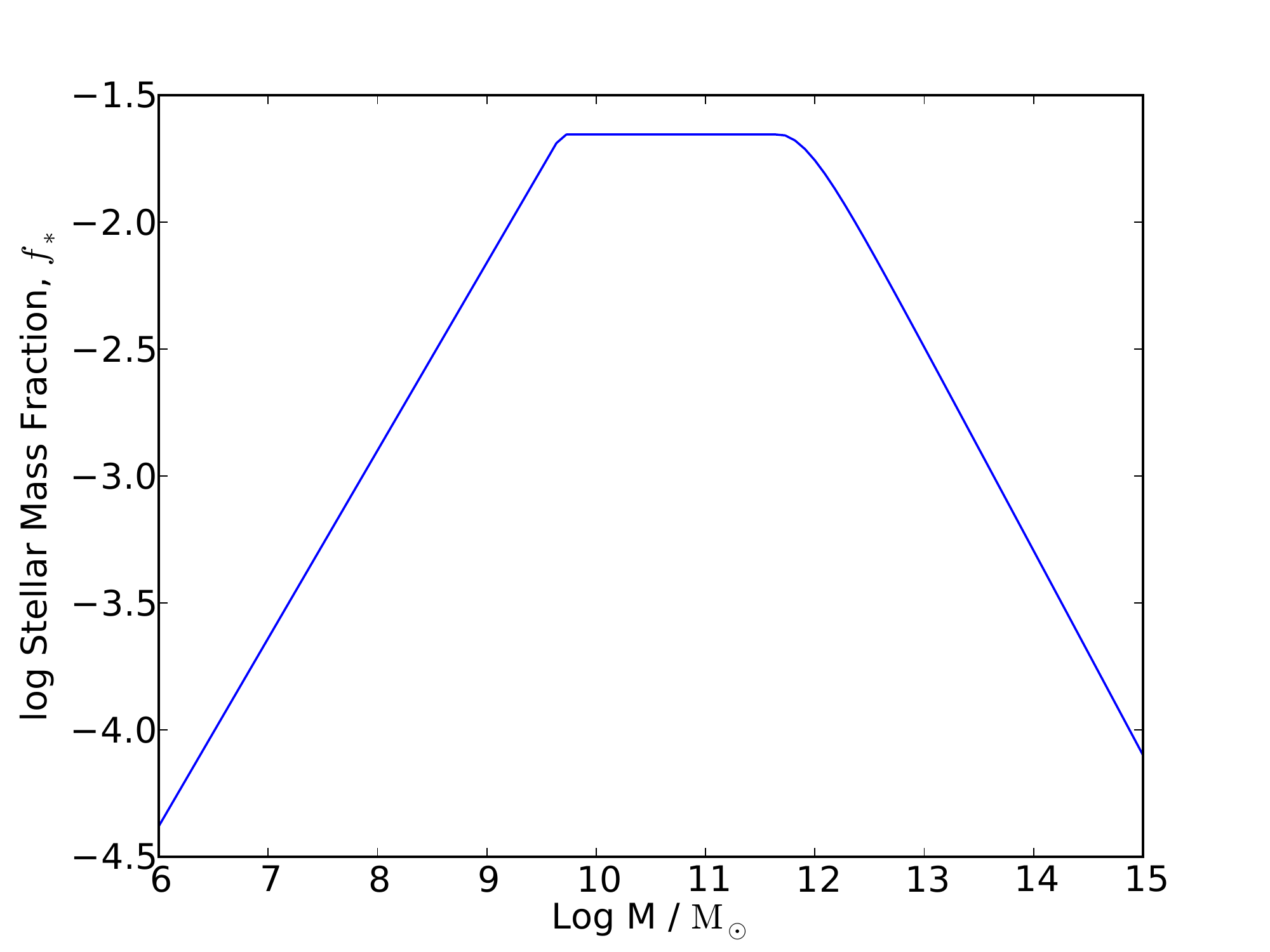}
	\caption{Stellar mass fraction as a function of host halo mass. The lower mass end power law is taken from \citet{OShea2015} while the high mass end is taken from \citet{Behroozi2013} at z = 6. We approximate the stellar mass fraction in the central mass ranges to be equivalent to the peaks of both ends to maintain continuity.}
	\label{fig:fstar}
\end{figure}

To determine the stellar mass fraction, we use a combination of fits taken from \citet{OShea2015} and \citet{Behroozi2013}. From the former, valid for the range $\mathrm{log}\ M_{\rm{vir}}/\msun < 10$, we have 

\begin{equation}
\label{eq:oshea}
f_* (M_{\rm{vir}})  = 1.26\ \rm{x}\ 10^{-3} \left( \frac{M_{\rm{vir}}}{10^8 \msun} \right)^{0.74}  \\ 
\end{equation}
fitted using data from the Renaissance Simulations that focus on galaxy formation during the EoR. These simulations have found that galaxy properties during EoR are largely independent of redshift \citep{Xu2016}. From the latter, valid for the range $\mathrm{log}\ M_{\rm{vir}} \geq 10$, we have

\begin{equation}
\label{eq:behroozi}
\mathrm{log}\ f_* (M_{\rm{vir}},z=6)  =   \mathrm{log}\ (\epsilon M_1) +f \left( \mathrm{log}\ \left(\frac{M_{\rm{vir}}}{M_1}\right)\right) - f(0) - \mathrm{log}\ M_{\rm{vir}}
\end{equation}
where $\epsilon$, $M_1$, and the function $f$ are heavily involved parameters.The exact details of this parameterization can be found in \citet{Behroozi2013}. We take this fit at only z = 6 and apply for all redshifts to maintain continuity for all mass ranges. Figure \ref{fig:fstar} shows the combined fits of $f_{*}$ at various redshifts. In order to remove discontinuities in combining the two fits, we extrapolate Eq. \ref{eq:oshea} until $f_* = 0.022$, or equal to the maximum of Eq. \ref{eq:behroozi} at z = 6. Then we assume a constant $f_*$ in the range between the two fits to connect them continuously. This imposed ceiling is largely consistent with the results from high-redshift numerical simulations which show a maximum stellar fraction \citep{Hopkins2014, Kimm2014, Schaye2014, Pawlik2016}.

For galaxies, we take the duty cycle to be the halo occupation fraction, $f_{\rm{occ}}$. This parameter takes into account for the fact that not every halo has a stellar population that provides ionising photons. Thus, it dampens the contribution from young low-mass halos which have bursty star formation periods. This relation is given by 

\begin{equation}
f_{\rm{occ}}(M_{\mathrm{h}}) = \left[ 1+ \left( 2^{\alpha/3} - 1 \right) \left( \frac{M_\mathrm{h}}{M_\mathrm{c}}\right)^{-\alpha}\right]^{-3/\alpha}
\end{equation}
taken from \citet{OShea2015} based on the form from \citet{Okamoto2008} where $\alpha = 1.5$ and $M_\mathrm{c} = 6.0$ x $10^7 \msun$. This function exponentially drops off below 1 for masses below the characteristic mass, $M_\mathrm{c}$. Above this mass, the fraction quickly approaches unity implying every halo contains ionising sources. Finally, we take $N_{2,\rm{\gamma / b}}$ to be a constant 4000 photons per baryon \citep{Schaerer2003}.

\subsubsection{Recombinations}

Furthermore, we introduce a mean recombination number per baryon as 

\begin{equation}
n_{\rm{rec}} = C(z)\ t_{\rm{H,0}}\ \alpha_{\mathrm{B}}\ \bar{n}_{\rm{H,0}}\ (1+z)^{3/2}
\end{equation}
where $t_{\rm{H,0}}$ is the Hubble time at the present day, $\bar{n}_{\rm{H,0}}$ is the mean hydrogen number density at the present day, and $\alpha_\mathrm{B}$ is the case B recombination coefficient at $10^4$ K which is taken as 2.6 x $10^{-13} \rm{cm}^{3}\ \mathrm{s}^{-1}$. As our treatment of the recombination number is a global value that only depends on redshift, and not on the halo mass, we can safely evaluate it outside the integral. We also include the clumping factor given by 

\begin{equation}
C(z) = \begin{cases}
1 + \mathrm{exp}(-0.28z + 3.50) & z \geq 10 \\
3.2 & z < 10
\end{cases}
\end{equation}
taken from \citet{Pawlik2008} to account for the boosted recombination rates in a clumpy IGM. An increased recombination rate requires an increased number of photons to keep the IGM reionised which has the effect of dampening $\zeta$ overall.  This is in contrast to the method of \citet{Sobacchi2014} where the recombination rate was calculated in each cell to produce the time-integrated number of recombinations per baryon to adjust $\zeta$.

\subsubsection{Putting it Together}

\begin{figure}
	\includegraphics[width=\columnwidth]{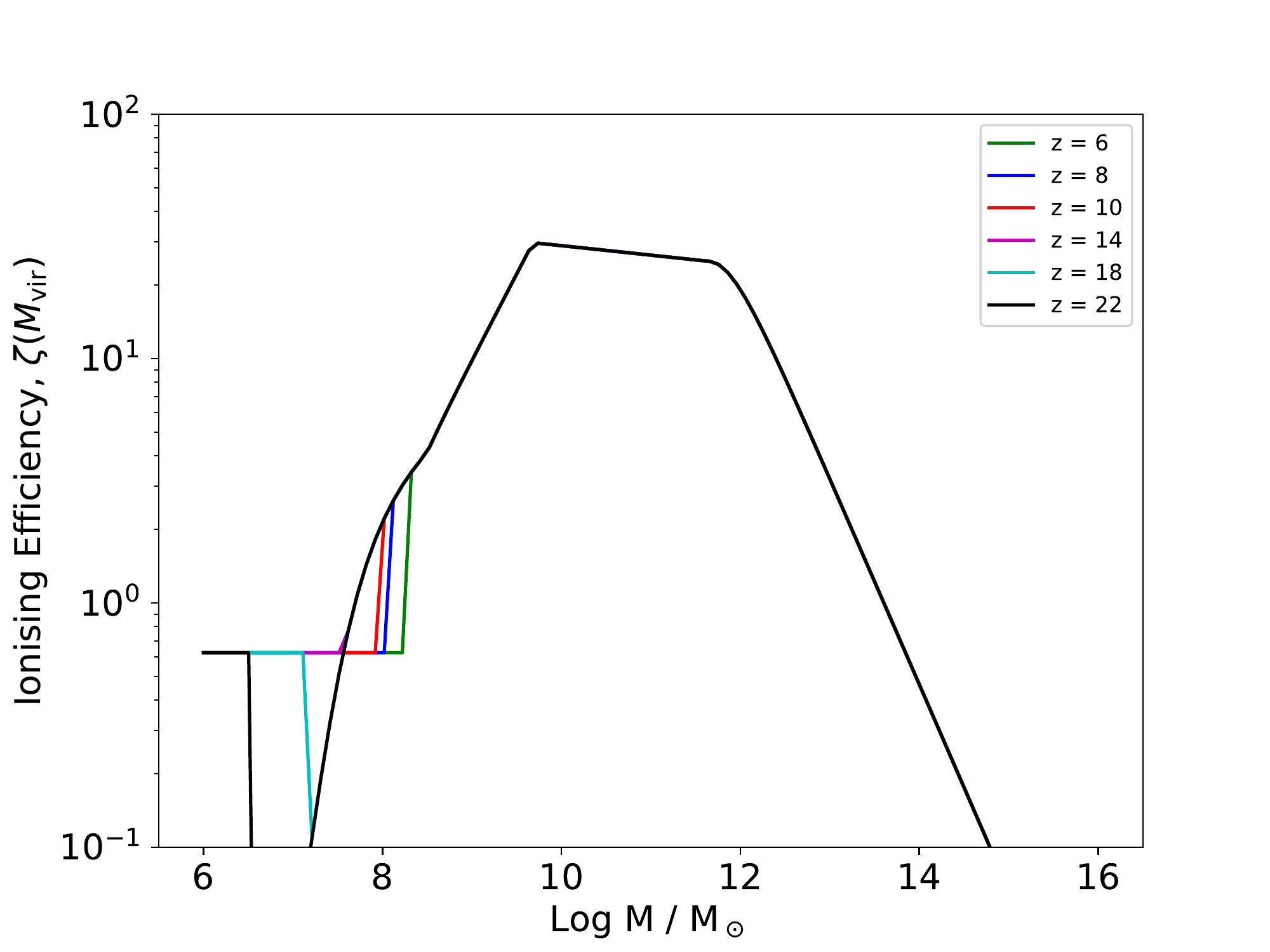}
	\caption{Ionising efficiency, $\zeta$, as a function of host halo mass at various redshifts. For masses $M_{\rm{min}} \leq  M_{\rm{vir}} < M_{\rm{filter}}$, we assume Pop III stars are the dominant ioinisation sources and assume a constant $\zeta$. Both $M_{\rm{min}}$ and $M_{\rm{filter}}$ evolve with redshift and the latter acts a moving threshold between mini-halos and galaxies. For galaxies, we take the distribution of ionising efficiencies as a function of the halo mass to be independent of redshift. }
	\label{fig:zetamass}
\end{figure}

At this point, we now have an ionization efficiency as a function of halo mass at different redshifts as shown in Figure \ref{fig:zetamass}. The biggest contribution in the range $9 \leq \mathrm{log}\ M_{\rm{vir}}/\msun< 12$ is due to the peaking of $f_*$. These galaxies have large $f_*$ while still having \fesc > 0.1 and thus provide the largest fraction of ionising photons. For $\mathrm{log}\ M_{\rm{vir}}/\msun > 12$, star formation becomes inefficient represented by a steep decline in $\zeta$. At the lower mass end below $\mathrm{log}\ M_{\rm{vir}}/\msun < 8$, the star forming halo occupation fraction greatly depresses $\zeta$. While we assume redshift-independent star formation parameters for $\zeta$ above the filtering mass, we see different values of $\zeta$ for different redshifts as a result of the clumping factor's redshift dependence.

In order for 21cmFAST to accurately calculate the collapse mass fraction, $f_{\rm{coll}}$, we then include the ionization efficiency term into the integral over the conditional mass function. The values of $\zeta$ are tabulated for a number of redshifts and then bi-linearly interpolated between different mass and and redshift values in the numerical integration. 

Given these distributions, we take a weighted average to get a single $\zeta$ value for a given redshift. First, we take the halo mass function (HMF) at a given redshift. We use the following form
\begin{equation}
\frac{dn}{dM} = f(\sigma) \frac{\bar{\rho}_{m}}{M} \frac{d\ \mathrm{ln}\ \sigma^{-1}}{dM}
\end{equation}
taken from \citet{Tinker2008} calculated using the python software package {\sc{Rabacus}} \citep{Altay2015}. We then normalize the HMF at $M_{\rm{min}} (z)$ to be 1. Now we can define $n_{\rm{frac}}$ to be the fraction of halos at a mass range between $M$ and $M+dm$ by taking the difference of the normalized HMF at those values. 

We can then take the integrated average of $\zeta$ weighted by halo number density fraction and halo mass as

\begin{equation}
\zeta (z) = \frac{\int_{M_{\rm{min}}}^{M_{\rm{max}}} \zeta (M_{\mathrm{h}})\ n_{\rm{frac}}\ M_{\mathrm{h}}\ dm}{\int_{M_{\rm{min}}}^{M_{\rm{max}}} n_{\rm{frac}}\ M_{\mathrm{h}}\ dm} 
\end{equation}
where $n_{\rm{frac}}$ is the fraction of halos within a mass range between $M_\mathrm{h}$ and $M_\mathrm{h} + dm$. We find take $M_{\rm{max}} = 10^{15} \msun$ to consider the full range of halo masses. This integral is similar to that introduced in \citet{Furlanetto2005}, where they took $\zeta$ to have a power-law dependence on the host halo mass instead.

Finally, we can then calculate a $\zeta$ for any given redshift. We fit a polynomial to $\zeta$ with the functional form given by 

\begin{equation}
\label{eq:fit}
\zeta (z) = A + B z^2 + C z^2 + D z^3 + E z^4
\end{equation}
where the coefficients are shown in Table \ref{tab:coeff}. At high redshifts, $\zeta$ remains mostly constant. This is because at these redshifts, the vast majority of ionising sources are mini-halos whose ionising efficiencies we have taken to be a constant value significantly lower than that of galaxies. These smaller objects form smaller \hii{} regions and thus cover only a small volume fraction of the total universe. In contrast, the general trend shows an exponential increase in the ionising efficiency at lower redshifts. Recall that the ionising efficiencies peak in the range 9.5 < log $M_{\rm{vir}}/\msun$  < 12. At these lower redshifts, the number of halos available to produce ionising photons at these mass ranges continually increases as halos merge to form larger structures which results in the boosted $\zeta$. These galaxies with large $f_*$ provide the bulk of ionising photon budget necessary for reionisation. We stress that Eq. \ref{eq:fit} is only valid for the range $5 <$ z $< 25$ as all the parameters have been calibrated from high-redshift simulations.

\begin{table}
  \caption{Coefficients for fits of $\zeta$}
  \label{tab:coeff}
  \begin{tabular}{lccccc}
   \hline
   Model & A  & B & C & D & E \\
   \hline
    Mean & 20.96 &  -4.871 & 0.425  & -1.622 x $10^{-2}$ & 2.265 x $10^{-4}$ \\
    Lo & 10.18 &  -2.376 & 0.207  & -7.920 x $10^{-3}$ & 1.111 x $10^{-4}$ \\
    Hi & 35.95 &  -8.299 & 0.725  & -2.756 x $10^{-2}$ & 3.820 x $10^{-4}$ \\
    \hline
  \end{tabular}
\end{table}

\subsubsection{Quantifying the Uncertainties}
\label{sec:uncertainties}

In order to consider the full range of values given the large uncertainties in certain parameters, we calculated the upper and lower limits to $\zeta$ as a function of redshift. Table \ref{tab:uncertainty} shows the list of parameters that we have chosen to vary along with the range. The greatest variances are in \fesc\ reported by \citet{Wise2014} and \citet{Kimm2014} which dominate the uncertainties for Pop III stars. For $f_\mathrm{d}$, we assume the same lifetime for Pop III stars and only vary the recovery times as reported by \citet{Muratov2012} and \citet{Jeon2014}. For the galactic $f_*$, we take the average variances found in \citet{Behroozi2013}. All other parameters not listed in the table remain as their original definitions.

\begin{table}
  \caption{Varied parameters and their values}
  \label{tab:uncertainty}
  \begin{tabular}{lccc}
   \hline
   Parameter & Mean\ Value & Lo\ Value & Hi\ Value \\
   \hline
    Pop III \fesc  & 0.6 & 0.05 & 0.9 \\
    $f_\mathrm{d}$ & $\frac{5\ \rm{Myr}}{30\ \rm{Myr}}$ & $\frac{5\ \rm{Myr}}{100\ \rm{Myr}}$ &  $\frac{5\ \rm{Myr}}{10\ \rm{Myr}}$\\
    Galaxy \fesc & \fesc ($M_\mathrm{h}$) & \fesc ($M_\mathrm{h}$) x 0.7 & \fesc ($M_\mathrm{h}$) x 1.3 \\
    Galaxy $f_*$ & $f_*(M_{\mathrm{h}},z)$ & $f_*(M_{\mathrm{h}},z)$ x 0.7 & $f_*(M_{\mathrm{h}},z)$ x 1.3 \\
    \hline
  \end{tabular}
\end{table}

These values are used to  produce the ionisation fields for the upper, lower, and standard values of $\zeta$. They provide a first order approximation to the possible distribution of $\zeta$ values. For both the lower and upper limits, we take a polynomial fit of the same form as Eq. \ref{eq:fit} to calculate $\zeta$. The coefficients for the resulting fits are found in Table \ref{tab:coeff}.

\begin{figure}
	\includegraphics[width=\columnwidth]{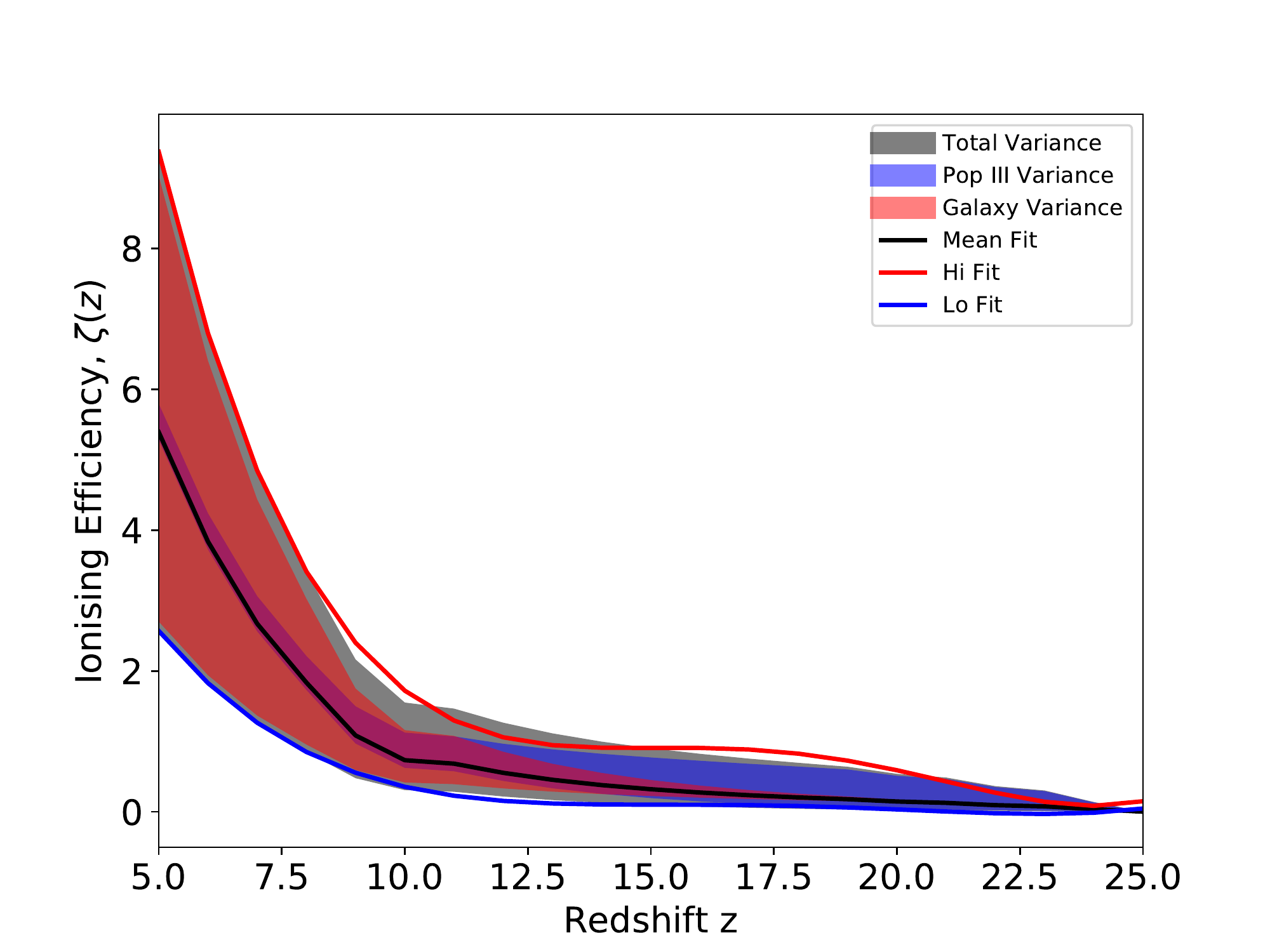}
	\caption{Integrated ionising efficiency $\zeta$ as a function of redshift including the spread using both low and high limit values for the various parameters in Eq. \ref{eq:zeta}. The blue and red shaded regions show the spread of $\zeta$ due to Pop III and galaxies respectively, while the grey shows the total spread due to the combined variance. The lines also show the polynomial fit to each of the Lo (blue), Hi (red), and Mean (black) values of $\zeta$. }
	\label{fig:var}
\end{figure}

Figure \ref{fig:var} shows the corresponding variances in the $\zeta$ function. The blue shaded region shows the resulting variance due to Pop III parameters while the red region shows it for galaxies. The effective combined range of values are represented by the grey area. At high redshifts (i.e. z > 15), the spread is entirely blue indicating only the mini-halos contribute significantly to the photon budget. As structure formation continues, the galactic contribution dominates after z < 10. This is expected as the Pop III star formation rate plateaus as their own formation results in the metal-enrichment of their surroundings suppressing further Pop III formation \citep{Wise2012, Xu2016a} . Instead, these mini-halos merge together to assemble galaxies with greater star formation rates and larger collapsed structures. Once reionisation is fully underway, the galactic contribution increases exponentially which also increases the spread of uncertainties at lower redshifts \citep{Sharma2016}.

\section{Results}
\label{sec:Results}

We run a total of four simulations from the same cosmological initial conditions at z = 300 each with varying $\zeta$ . The high resolution density grid is sampled by $2048^3$ cells which is smoothed over a $1024^3$ grid to produce the ionisation field. We produce 50 snapshots equally spaced in time starting from z = 25 down to z = 6 to produce the entire ionisation history. The fiducial case takes the three parameter model from \citet{Greig2016} consisting of $T_{\rm{vir}}^{\rm{min}}$, the minimum virial temperature hosting ionising sources, $\zeta$, the ionisation efficiency, and $R_{\rm{mfp}}$, the maximum horizon for ionising photons which defines the maximum filtering scale. From their results, we take the best fit values for each of the parameters which are $T_{\rm{vir}}^{\rm{min}} = 10^5$ K, $\zeta$ = 50, and $R_{\rm{mfp}} = 20$ Mpc. 

In comparison with the fiducial case, we run a total of three simulations with varying values of $\zeta$ as a function of redshift. In each of the runs, rather than taking the minimum virial temperature as a proxy for the minimum mass of ionising halos, we use the 
minimum mass calculated by Eq. \ref{eq:mmin}. The three runs are then the Mean, Hi, and Lo cases which represent the base fit to $\zeta$ and its upper and lower variance values with their fits given in Table \ref{tab:coeff}. We keep the same maximum horizon as $R_{\rm{mfp}} = 20$ Mpc as \citet{Sobacchi2014} and \citet{Greig2016} have shown that the resulting ionisation fields are largely insensitive to the choice. 

\subsection{Reionisation Histories}

\begin{figure}
	\includegraphics[width=\columnwidth]{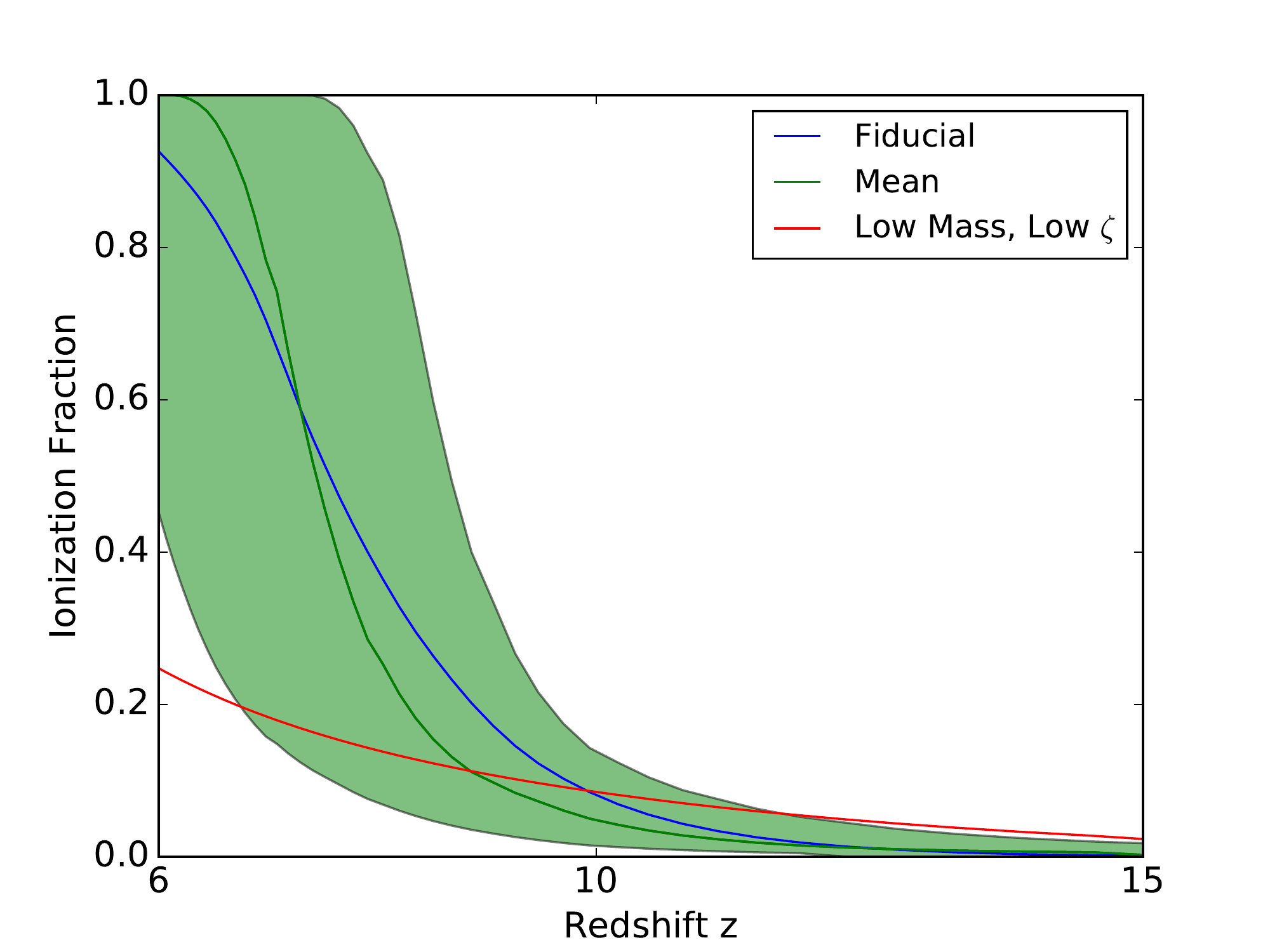}
	\caption{Ionisation histories of all four runs. The blue line represents the fiducial run using the three parameter model from \citet{Greig2016}, while the red also shows a three parameter model run but with a low mass and low zeta. The green line shows the Mean value run using our new parameterization. The shaded region shows the spread in ionisation histories using our Lo and Hi value parameterizations.  }
	\label{fig:hist}
\end{figure}

Figure \ref{fig:hist} shows the ionisation histories calculated from each of the runs. We define the start of reionisation, $z_{\rm{start}}$, to be when the ionised fraction, $x(z)$, is at 10$\%$. Similarly, the end, $z_{\rm{end}}$, is when $x(z) = 99\%$. The blue line shows the fiducial case which does not quite end up fully ionised at the end at z = 6.0, while $z_{\rm{start}} = 9.8$. The green line shows the Mean case which also has $z_{\rm{start}} = 9.0$ and $z_{\rm{end}} = 6.0$. The fiducial model has a much steeper rise at $z_{\rm{start}}$ while the Mean case shows a gradual rise in the ionised fraction. In the former, as only halos with $T_{\rm{vir}} > 10^5$ K are considered, there is a more abrupt increase in the ionised fraction as these halos do not exist in large numbers until lower redshifts. In the latter, as mini-halos begin forming early on at high redshifts, there is a gradual increase in the ionised fraction as Pop III stars continually add on to the photon budget. Moreover, since the value of $\zeta$ is relatively sensitive to Pop III parameters at $z \sim 10$, mini-halos must play a role in determining the exact starting point of reionisation. Once reionisation is underway, the Mean case shows a steeper increase in the ionised fraction resulting in a slightly earlier end to the EoR. This is mostly due to the steep incline in $\zeta$ at these low redshifts corresponding to the presence of bigger halos with large $f_*$ emitting a significant amount of ionising photons. The ionisation histories are relatively consistent with each other only differing by a maximum of 23 percent at lower redshifts. Overall, we find the fiducial model to be a reasonable approximation to our much more involved parametrization.

For comparison, we also run a three parameter model with a $T_{\rm{vir}}^{\rm{min}} = 5000$ K, $\zeta$ = 3 and the resulting ionization history is also shown in Figure \ref{fig:hist} in red. Given this lower minimum mass threshold, we chose a $\zeta$ such that the onset of reionisation is close to that of the other runs. However, because $\zeta$ in this particular model is a constant over redshifts, it is not sufficient to ionise the box by the end of the run with only 25$\%$ ionised at z = 6. Otherwise, increasing the $\zeta$ to fully ionise the box results in a significantly earlier reionisation starting redshift. This demonstrates the need for non-uniform $\zeta$ when attempting to incorporate the effects of low-mass halos. 

The shaded region in green shows the spread in histories where the edges represent the Hi and Lo value cases. The Hi value case has $z_{\rm{start}} = 10.5$ and $z_{\rm{end}} = 7.3$ while the Lo value case has $z_{\rm{start}} = 7.5$ and only reaches x = 0.55 at z = 6. Given the constraint that the universe is mostly ionised by z = 6, much of the lower spread in histories is effectively ruled out. This broadly constrains our parameters, in particular $\fesc$ and $f_*$ for galactic populations. However, even considering just the Hi case, there is a broad range of $z_{\rm{start}}$ as the large mini-halo population quickly drives up the ionised fraction to the threshold fairly early on.

We also calculate the optical depth due to Thomson scattering for each of the runs by 

\begin{equation}
\tau_e = \int_{0}^{\infty} dz \frac{c(1+z)^2}{H(z)} x(z) \sigma_\mathrm{T} \bar{n_\mathrm{H}}(1+\eta Y/4X)
\end{equation}
where H(z) is the Hubble parameter, $x(z)$ is the ionised fraction of hydrogen, $\sigma_\mathrm{T}$ is the Thomson cross-section, and $X$ and $Y = 1 - X$ are the hydrogen and helium number fractions respectively. We also assume that helium is singly ionised ($\eta$ = 1) at z > 3 and doubly ionised at later times ($\eta$ = 2). 

The fiducial case produces $\tau_e$ = 0.0567 while the Mean case has $\tau_e$ = $0.0557_{-0.0144}^{+0.0146}$, where the Hi and Lo case $\tau_e$ are represented as uncertainties. Taking the estimated value from the \textit{Planck} 2016 intermediate results of $\tau_e$ = 0.0596 $\pm$ 0.0089 \citep{Planck2016a}, we see that our value is still well within the margin of error.

\subsection{Bubble Size Distributions}

In order to further characterize the differences between our models, we generate ionised bubble size distributions. These distributions tell us about the morphology of reionisation as the \hii{} regions grow and expand. We use the same methodology found in \citet{Mesinger2007} to maintain consistency in generating the distributions. First, we smooth out the ionisation field and remove the partial ionisation values by setting a threshold. We choose this threshold to be 0.5. We then choose an ionised cell and a direction vector at random and measure the distance to the nearest neutral cell. We repeat the process $10^7$ times to get a 
distribution. This method has been shown to be a good approximation to getting more accurate distributions \citep{Lin2015}.

\begin{figure}
	\includegraphics[width=\columnwidth]{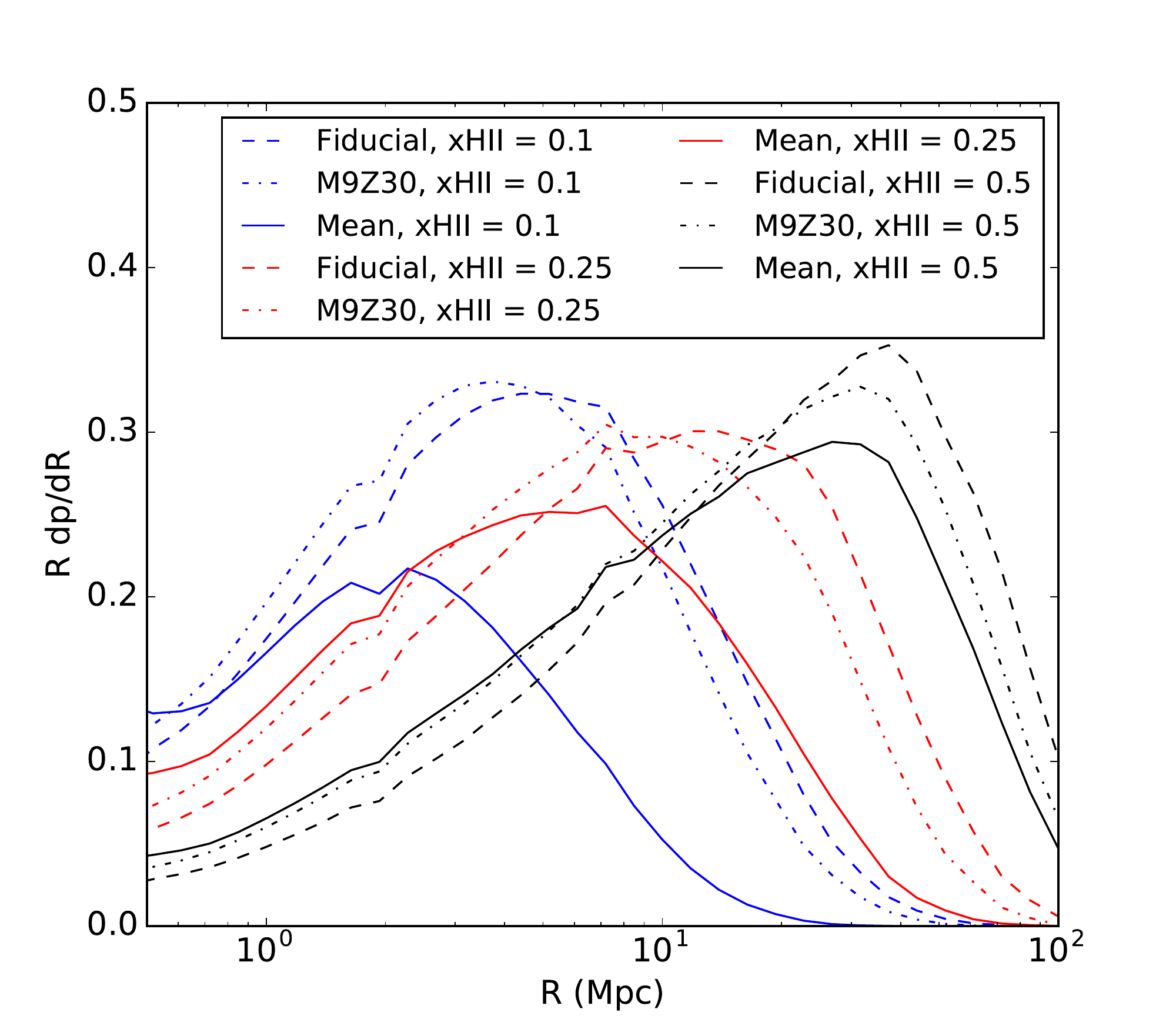}
	\caption{Bubble size distributions at $x \approx 0.1$ (blue), $x \approx 0.25$ (red), $x \approx 0.5$ (black)  . The Mean case is represented with dashed lines, the Fiducial with solid lines, and the M9Z30 run is shown in dash-dot lines. The dip and peak at R $\sim$ 2 Mpc and $\sim$ 8 Mpc are artifacts from the numerical bubble size distribution calculation.}
	\label{fig:compare}
\end{figure}

Figure \ref{fig:compare} shows the generated bubble size distributions for the Mean and Fiducial cases at different ionisation fractions. For additional comparison, we also include a three parameter run with $M_{\rm{min}} = 10^9 \msun$ , $\zeta$ = 30, labeled as M9Z30. At lower ionisation fractions, we see that the Mean case has a lower characteristic size compared to either of the three parameter runs. This can be understood as due to the presence of mini-halos at this early phase which make up a larger fraction of ionised cells in our model driving the peak down. These mini-halos have small ionising photon luminosities, which would correspond to smaller \hii{} regions. As reionisation progresses, however, we see that the size distribution of the Mean run begins to converge to the M9Z30 solutions. At lower redshifts, the reionisation topology is driven by the larger halos as their numbers begin to increase. Thus, the contribution from mini-halos is greatly suppressed at higher ionisation fractions, with some minimal contributions at smaller scales. Also at high $x$, the characteristic size quickly approaches the size of the box in both cases as expected. 

Recently \citet{Paranjape2014} showed that a correction to remove the correlation in the random walk introduced by the smoothing filter can result in a significant increase in the characteristic sizes. We expect a similar impact should the correction be included. 

\section{Discussion and Summary}
\label{sec:disc}

The greatest strength in our parameterization is that the assumed values are those constrained by the latest numerical simulations including full range of physical process including radiative and supernova feedback effects. This enables us to consider the full range of mass scales rather than assuming a single ionising efficiency for  all halos. However, one large downside to this particular semi-numerical treatise is that the method is still fundamentally a single parameter model that only depends on the collapse mass fraction. There is no consideration of the environment that the sources live in, whether it is in a ionised region or not, and we take a relatively crude average over all halos in different environments to get a single efficiency coefficient. This may result in an improper weighting of $\zeta$. This is a problem that is well treated in contrast in full radiation hydrodynamics simulations, which is now starting to be computationally feasible at large scales \citep{Ocvirk2015}. Moreover, another drawback is that this methodology is not entirely self-consistent as the assumed LW background and filtering masses are taken empirically rather than calculated on the fly.

A number of instruments will be coming online within the next several years to help put tighter constraints on models of reionisation. 21 cm interferometry performed by the Square Kilometer Array (SKA)\footnote{http://www.skatelescope.org/} and the Hydrogen Epoch of Reionization Array (HERA)\footnote{http://reionization.org} will produce accurate mapping of the morphology of the reionisation process. Moreover, the James Webb Space Telescope should extend the current limits to the luminosity function of galaxies constraining parameters such as the stellar mass fraction. With these observations, we expect that our models can be utilized to study the onset of reionisation.

In this work, we extended the semi-numeric simulation code {\sc{21cmFAST}} to include a redshift-dependent minimum mass threshold for ionising source containing halos, $M_{\rm{min}}$, as well as a mass- and redshift-dependent ionising efficiency, $\zeta$. Our model produces reionisation histories that have subtle differences in comparison with the default model implemented in {\sc{21cmFAST}} while still being broadly consistent with the constraints from \textit{Planck}. Moreover, we find significant differences in the bubble size distribution due to the presence of mini-halos which drive the characteristic scales down. We find that our model broadly constrains the minimum ionising efficiency contribution from galaxies while mini-halos only contribute near the beginning of reionisation, having no significant impact after z = 10.

\section*{Acknowledgements}

We thank Andrei Mesinger for making \textsc{21cmFAST} publicly
available and an anonymous referee for helpful comments and
suggestions that improved this paper.  This research was supported by
National Science Foundation (NSF) grants AST-1333360 and AST-1614333,
NASA grant NNX17AG23G, and Hubble theory grants HST-AR-13895 and
HST-AR-14326.





\bibliographystyle{mnras}
\bibliography{library} 




%


\bsp	
\label{lastpage}
\end{document}